%% file: main.tex
\RequirePackage{silence}
\documentclass[aps,prl, reprint,
dvipsnames]{revtex4-1}
\pdfoutput=1
\usepackage{blindtext}

\usepackage{graphicx}
\usepackage{dcolumn}
\usepackage{bm}
\usepackage{amssymb}
\usepackage{color}
\usepackage[T1]{fontenc} 
\usepackage{mathrsfs,amsfonts,dsfont}
\usepackage{nccmath}
\usepackage{amstext}
\usepackage{mathtools}
\usepackage{tablefootnote}
\usepackage[inline]{enumitem}
\graphicspath{{graphics/}}
\usepackage[english]{babel}

\usepackage[table]{xcolor}

\usepackage{wrapfig}
\usepackage{lipsum} 
\usepackage{blindtext}

\usepackage{braket}
\usepackage{bbm} 
\usepackage[normalem]{ulem}

\usepackage[nolist,withpage]{acronym}
\makeatletter
\AtBeginDocument{%
  \renewcommand*{\AC@hyperlink}[2]{%
    \begingroup
      \hypersetup{hidelinks}%
      \hyperlink{#1}{#2}%
    \endgroup
  }%
}
\makeatother

\usepackage[colorlinks=true,citecolor=blue,linkcolor=blue]
{hyperref}

\usepackage{amsthm}

\newtheorem*{mainresult}{Main result}

 \hypersetup{pdftitle = {Quantum correlations cannot be reproduced with a finite number of measurements in any no-signaling theory},
       pdfauthor = {Lucas Tendick},
       pdfsubject = {Quantum information theory}, 
       pdfkeywords = {quantum, assemblage,  theory, incompatibility, 
       causality, genuine, causal theory, no-signaling, non-signaling, 
       quantum measurements, nonlocality, Bell, post-quantum,
       alternatives to quantum theory, genuine, genuinely incompatible, 
       finite number of measurements, infinite number of measurements
              }
      }

\begin{document}
\title{Quantum correlations cannot be reproduced with a finite number of measurements in any no-signaling theory}
\author{Lucas Tendick}
\email{lucas-amadeus.tendick@inria.fr}
\affiliation{Inria, Université Paris-Saclay Palaiseau, France}
\affiliation{CPHT, CNRS, Ecole Polytechnique, Institut Polytechnique de Paris, Palaiseau, France}
\affiliation{LIX, CNRS, Ecole Polytechnique, Institut Polytechnique de Paris, Palaiseau, France}

\begin{abstract}
We show, for any finite $n \geq 2$, that there exist quantum correlations obtained from performing $n$ dichotomic quantum measurements in a bipartite Bell scenario, which cannot be reproduced by mixtures of measurement devices with at most $(n-1)$ incompatible measurements across different partitions in any no-signaling theory. That is, it requires any no-signaling theory an unbounded number of measurements to reproduce the predictions of quantum theory. We prove our results by showing that there exist linear Bell inequalities that have to be obeyed by any no-signaling theory involving only $(n-1)$-wise incompatible measurements and show explicitly how these can be violated in quantum theory. Finally, we discuss the relation of our work to previous works ruling out alternatives to quantum theory with some kind of bounded degree of freedom and consider the experimental verifiability of our results.       
\end{abstract}

\maketitle

\indent \textit{Introduction.}\textemdash The goal of theoretical physics is to devise a model of nature that accurately explains past observations and allows for experimentally verifiable predictions. Quantum theory is the most successful and accurately tested description of nature that exists in physics \cite{PhysRevLett.100.120801, Sailer2022}. Despite $100$ years of proving its success, the effectiveness of quantum theory is still puzzling from a theory-building perspective \cite{Wheeler1989-WHEIPQ, QuantumTheory2016}. In its essence, quantum theory is a set of mathematical rules that lead to observable predictions. \\
\indent Notably, several works indeed derive the mathematical formalism from first principles \cite{quant-ph/0101012, PhysRevA.84.012311, Barnum2014, PhysRevA.100.032120, 0911.0695, Masanes2011}. However, the search for the precise physical principles \cite{POPESCU1992293,Pawowski2009,Fritz2013,PhysRevLett.96.250401,PhysRevLett.99.180502,Navascus2009,Navascus2015} that allow for a derivation
of (the limits of) quantum theory purely from observations, i.e., the input-output statistics of local measurements in a Bell scenario with uncharacterized devices (black-boxes) is still ongoing.  \\
\indent Initially devised to rule out the possibility of theories based on the premise of \emph{local hidden variables} \cite{EPR_paper}, which has been demonstrated in many groundbreaking experiments \cite{PhysRevLett.119.010402, PhysRevLett.115.250401, Hensen2015, PhysRevLett.115.250402}, Bell's theorem \cite{Bell_seminal} and Bell inequalities found many applications beyond their initial use \cite{Nonlocality_review,Tavakoli_2022}. The violation of a Bell inequality certifies the presence of entangled  systems \cite{RevModPhys.81.865, PhysRevLett.111.030501} and incompatible measurements \cite{RevModPhys.95.011003, Heinosaari2016,PhysRevResearch.3.023143, PhysRevLett.123.180401, PhysRevLett.48.291, PhysRevLett.103.230402, PhysRevA.93.052112}. Moreover, Bell inequality violations are a necessary prerequisite to self-test quantum systems \cite{Supic2020selftestingof}, to prove the security of cryptography protocols with minimal assumptions \cite{PhysRevLett.67.661,PhysRevLett.98.230501}, and to witness the generation of randomness \cite{https://doi.org/10.48550/arxiv.0911.3814, Pironio2010, Colbeck2011}. \\ 
\indent Crucially, entanglement and incompatibility are necessary requirements for the violation of a Bell inequality not only in quantum theory but also in much more general theories leading to probabilistic observations, so-called \acp{GPT} \cite{PhysRevLett.104.140404,Plvala2023,Banik2015,PhysRevA.94.042108}, which quantum theory is a special case of. Moreover, alternative theories of nature that are close to quantum theory \cite{Renou2021, PhysRevLett.127.200401,PhysRevLett.117.150401,PhysRevLett.88.170405, PhysRevLett.89.060401} can be ruled-out through the quantum violation of Bell-like inequalities restricting correlations obeyed by these theories. \\
\indent Judiciously chosen Bell scenarios have been used in the past to rule out no-signaling theories reproducing quantum correlations with only: a finite number of measurement outcomes \cite{PhysRevLett.117.150401}, a finite number of entangled parties \cite{PhysRevLett.88.170405, PhysRevLett.89.060401}, and sources of correlations of finite degree \cite{PhysRevLett.127.200401}. Hence, Bell-like inequalities can be used to prove that quantum correlations do not allow for a finite bound on many \emph{external parameters} (i.e., parameters that are not part of the mathematical theory but more of the Bell experiment itself). 
\begin{figure}[t]
\includegraphics[scale=0.34]{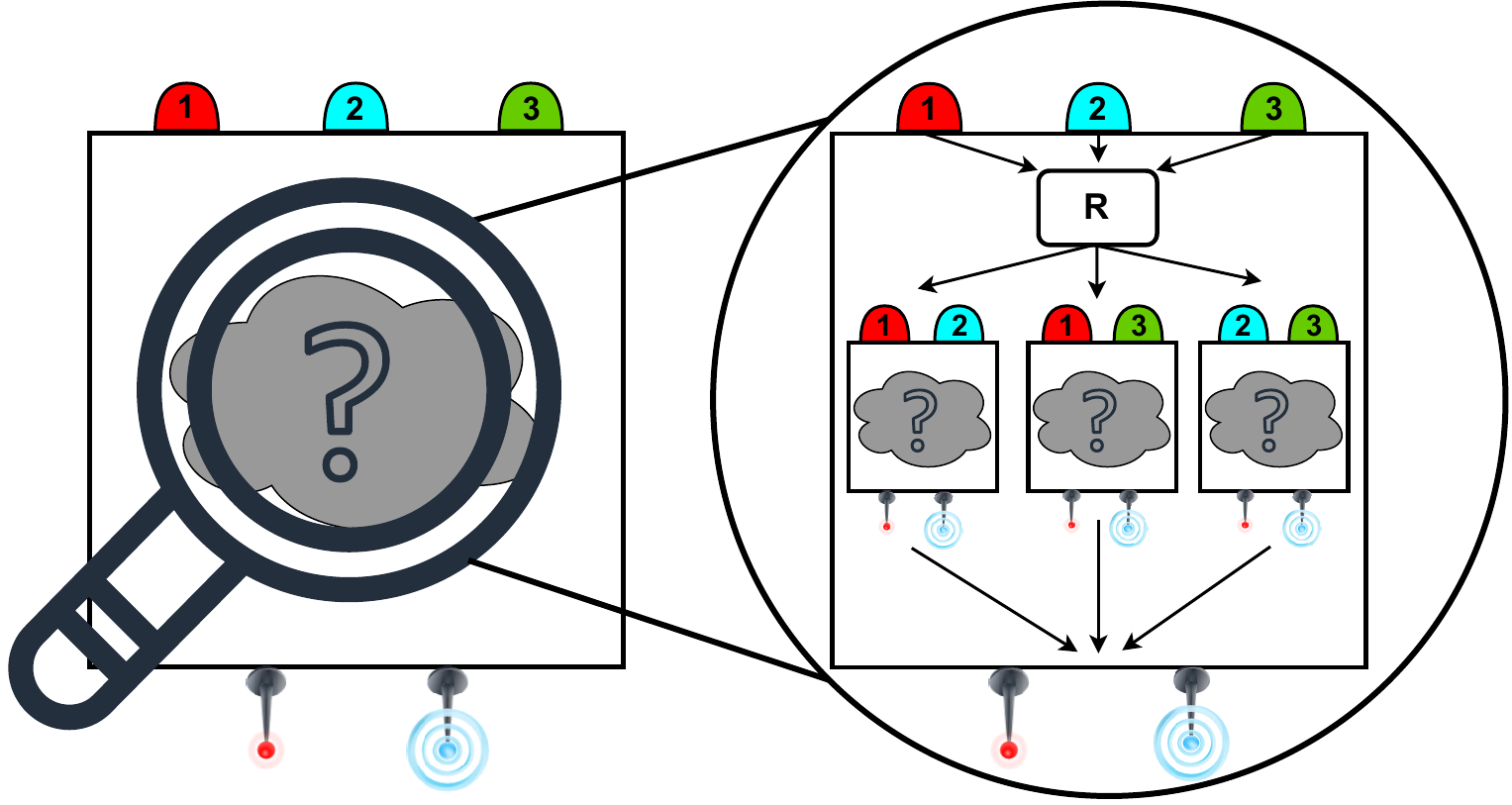}
 \caption{Graphical representation of our main question for the special case of $3$ measurements. Measurement devices are treated as black-boxes, i.e., we only observe the input-output statistics of the devices. Moreover, we only assume the devices to be restricted by the no-signaling principle and do not refer specifically to the quantum formalism. Just on the basis of the input-output statistics, we rule out that the inner workings of the measurement device include a randomization process $\mathrm{R}$ and measurement devices with fewer inputs to simulate the considered device.}
  \label{IntroductionScheme}
\end{figure}
 However, the question regarding the possible boundlessness of the number of measurements needed to describe quantum correlations remains unanswered. Surprisingly, recent progress \cite{PhysRevA.96.032104, PhysRevLett.123.180401} has been only able to show that quantum theory cannot be reproduced with just $2$ measurements (see also Figure \ref{IntroductionScheme}), leaving a general answer to the question: \emph{Can quantum theory possibly be reproduced by a (maybe very exotic) no-signaling theory that only requires a finite number of measurements?} widely open. \\
\indent In this work, we answer exactly this question in the negative. In particular, we show the existence of quantum correlations obtained from performing $n$ different dichotomic measurements on an bipartite entangled quantum state that cannot be reproduced by any theory  (classical, quantum, and beyond), which is compatible with the no-signaling principle involving only $(n-1)$ measurements for every finite $n$. Hence, every possible way to come up with a no-signaling theory reproducing quantum theory in which there is an (arbitrarily large but finite) upper limit on the number of different measurement settings, can be falsified. \\
\indent To prove our result, we consider the theory of measurement incompatibility, whose extension to recently studied incompatibility structures \cite{PhysRevA.89.052126, Heinosaari2008,Liang2011,PhysRevLett.123.180401,PhysRevLett.131.120202} is crucial to formally define what it means to perform genuinely $n$ different measurements. More precisely, we use that measurements which are genuinely $n$-wise incompatible, cannot be reduced to effectively $(n-1)$ measurements as convex combinations of assemblages that are at most $(n-1)$-wise incompatible. We then connect genuine $n$-wise incompatibility to the family of $M_{nn22}$ inequalities \cite{Brunner2006}, to show that a violation of these inequalities implies the existence of genuinely $n$-wise incompatible measurements in any no-signaling theory. We then show that any of the $M_{nn22}$ can be violated by quantum theory with a quantum state of local dimension $n$. Finally, we discuss possible experimental implementations of our findings.  \\
\indent \textit{Incompatible measurements in Bell experiments.}\textemdash Let us start by reviewing the notions of measurement incompatibility. In quantum theory, measurements are described by positive semidefinite operators $A_{a \vert x} \geq 0$ acting on a Hilbert space $\mathcal{H} \cong \mathds{C}^{d}$ of dimension $d$, such that the completeness relation $\sum_{a} A_{a \vert x} = \mathds{1}$, (where $\mathds{1}$ is the identity operator), is fulfilled. Here, $x$ denotes the label of a specific measurement and $a$ the label of the corresponding outcome to that measurement. We say a set of measurements (also known as assemblage) $\mathcal{A} = \lbrace \lbrace A_{a \vert x} \rbrace_a \rbrace_x$ is \emph{jointly measurable} if it can be simulated by performing a single measurement $ \lbrace G_{\lambda} \rbrace_{\lambda}$ (the so-called parent \ac{POVM}) and some classical post-processing via the conditional probabilities $ p(a \vert x, \lambda) $ such that 
\begin{align}
A_{a \vert x} = \sum_{\lambda} p(a \vert x, \lambda) G_{\lambda},\label{DefJM} 
\end{align}
holds. An assemblage that is not jointly measurable is called incompatible. Conceptually, joint measurability can be seen as the generalization of
commutativity of projective measurements to general quantum measurements \cite{Heinosaari2016,RevModPhys.95.011003}. \\ 
\indent To certify the incompatibility of measurements without characterizing the measurement device, one can rely on the device-independent (black-box) paradigm and the violation of Bell inequalities. Let us consider a bipartite Bell experiment in which two distant parties, Alice and Bob, perform one out of $n$ measurements in each round of the experiment labeled by $x$ and $y$, respectively. In each round, Alice obtains one out of two outcomes $a$, similarly Bob obtains outcome $b$. Treating their devices as black-boxes, all information about the experiment that is accessible to Alice and Bob are the input-output statistics, also called behavior or correlations, $\mathbf{P} = \lbrace p(ab \vert xy) \rbrace$. Just on the basis of the obtained behaviour $\mathbf{P}$ we want to prove that the considered black-box devices involve indeed incompatible measurements and hence cannot be treated as a single measurement. \\
\indent In the case of $n=2$ measurements we can do so by looking at the \ac{CHSH} inequality \cite{PhysRevLett.23.880} given by (in the notation of Collins-Gisin \cite{Collins2004})
\begin{align}
p(00 \vert 11) + p(00 \vert 12) &+ p(00 \vert 21) - p(00 \vert 22) \nonumber \\
&-p_A(0 \vert 1) - p_B(0 \vert 1) \leq 0. \label{CHSH}     
\end{align}
\indent By definition of being a Bell inequality, the \ac{CHSH} inequality has to be obeyed by all local (also known as classical) theories, i.e., by all theories where the correlations $\mathbf{P}$ could be explained via pre-shared randomness. However, it is well-known that quantum correlations obtained from performing quantum measurements on a shared bipartite quantum state $\rho$ such that $p(ab \vert xy) = \mathrm{Tr}[(A_{a \vert x} \otimes B_{b \vert y}) \rho]$ can violate the \ac{CHSH} inequality, provided judiciously chosen incompatible measurements $\mathcal{A}, \mathcal{B}$ are performed on an entangled state $\rho$. That is, the \ac{CHSH} inequality constitutes a device-independent certificate of entanglement and measurement incompatibility. While incompatibility is necessary for the violation of a Bell inequality, it is known that is generally not sufficient \cite{Bene2018,PhysRevA.97.012129}, however all incompatibility structures can be certified device-independently \cite{PhysRevA.110.L060201}. \\ 
\indent In their seminal work, Wolf et al. \cite{PhysRevLett.103.230402} prove that the violation of any Bell inequality (and in particular the CHSH inequality) not only implies the incompatibility of measurements in quantum theory, but in any theory that obeys the \emph{no-signaling principle}. The no-signaling principle states that the measurement devices of Alice and Bob cannot be used for superluminal signaling, which means Alice is oblivious of Bob's measurement setting and vice versa. More formally, the correlations $\mathbf{P}$ obey the no-signaling principle \cite{POPESCU1992293, Janotta2012}, if their marginal distributions $p_A(a \vert x)$ and $p_B(b \vert y)$ are well-defined and independent of the input of the other party, i.e., $p_A(a \vert x) = p_A(a \vert x y) = \sum_b p(ab \lvert xy)$
for all $a,x,$ and similarly for Bob's marginal. However, the general question of whether there always exist $n$ quantum measurements, for any $n\geq 2$, that cannot be reproduced by at most $(n-1)$ measurements in any no-signaling theory remains still open. \\
\indent \textit{Boundlessly many genuine incompatible measurements.}\textemdash 
Here, we answer this question affirmatively. To that end, we first have to define properly what it means to perform genuinely $n$ different measurements through the notion of genuinely $n$-wise incompatible measurements. While ordinary incompatibility according to Eq.~\eqref{DefJM} only guarantees that the assemblage $\mathcal{A}$ cannot be reduced to a single measurement, it does not guarantee that $n$ measurements cannot be reduced to fewer measurements in general. Using the notion of recently studied incompatibility structures, the authors in \cite{PhysRevLett.123.180401} define an assemblage to contain (at least) genuinely $n$-wise incompatible measurements if it cannot be implemented by probabilistically employing assemblages that contain at most genuinely $(n-1)$-wise incompatible measurements. That is, if the assemblage $\mathcal{A}$ cannot be decomposed such that 
\begin{align}
A_{a \vert x} = \sum_{(s,t \neq s) \in \mathcal{T}} p_{(s,t)}   F_{a \vert x}^{(s,t)}, \label{DefnGenuineIncomp}
\end{align}
where $\mathcal{F}^{(s,t)} \in \mathrm{JM}^{(s,t)}$ are assemblages, employed with probability $p_{(s,t)}$, in which the pair $(s,t \neq s)$ is jointly measurable. This makes each $\mathcal{F}^{(s,t)}$ contain at most genuinely $(n-1)$-wise incompatible measurements (as the pair $(s,t \neq s)$ is one effective measurement) and the set $ \mathcal{T}$ contains all tuples $(s,t \neq s)$ constructed from the integers $\lbrace 1, \cdots, n \rbrace$. \\
\indent The key observation of \cite{PhysRevLett.123.180401} is using the fact that an incompatibility structure like in Eq.~\eqref{DefnGenuineIncomp} results in restrictions on the observable behaviour $\mathbf{P}$ that then can be used in Bell scenarios to certify that certain measurements do not obey a given incompatibility structure. More precisely, assuming that Bob's measurements obey an incompatibility structure according to Eq.~\eqref{DefnGenuineIncomp} leads to a decomposition of the behaviour $\mathbf{P}$ of the form
\begin{align}
p(ab \vert xy) = \sum_{(s,t \neq s) \in \mathcal{T}} p_{(s,t)}  p^{(s,t)}(ab \vert xy), \label{DefnIncompDI}   
\end{align}
where $p^{(s,t)}(ab \vert xy)$ describes a behaviour $\mathbf{P}^{(s,t)}$ that is local once it is restricted to only the measurements $(s,t)$ on Bob's side. Now, crucially, disproving that a decomposition according to Eq.~\eqref{DefnIncompDI} exists while only assuming the no-signaling principle, constitutes a device-independent proof of genuinely $n$-wise incompatible measurements in any no-signaling theory. \\
\indent The numerical analysis for the case $n = 3$ in \cite{PhysRevLett.123.180401} reveals that while it is easy to use many well known inequalities like the $3$-\ac{CHSH} game \cite{PhysRevLett.121.180503}, the elegant \cite{quant-ph/0702021}, or the chained Bell inequality \cite{Braunstein1990} for the certification of genuinely $3$-wise measurements \emph{assuming quantum theory}, it is more difficult to rule out the possibility of a simulation by only $2$ measurements in general no-signaling theories. Essentially, there exists only one relevant Bell inequality, the $M_{3322}$ inequality for this task (see also  \cite{PhysRevA.96.032104}), which has also already been considered in \cite{quant-ph/0412109}. However, the $M_{3322}$ inequality only allows for a relatively small violation within quantum theory and employing numerical methods can only provide insights for small number of measurements $n$. Hence, despite providing the best current progress, the methods in \cite{PhysRevLett.121.180503} are of limited use for answering our main question. \\
\indent We show in the following that a detailed analytical analysis of why the $M_{3322}$ certifies genuinely $3$-wise incompatible measurements in any no-signaling theory leads to the key observation that enables us to prove our main result. More precisely, it enables us for every finite integer $n \geq 2$, to propose a Bell experiment, including $n$ dichotomic quantum measurements, that cannot be reproduced by any no-signaling theory using only $(n-1)$-wise compatible measurements. Let us denote the set of no-signaling correlations of at most genuinely $(n-1)$-wise incompatible measurements by $\mathrm{GNS}_{n-1}$. That is, $\mathbf{P} \in \mathrm{GNS}_{n-1}$ if a decomposition according to Eq.~\eqref{DefnIncompDI} exists. 
\begin{mainresult} 
For every $n \geq 2$ there exist quantum correlations $\mathbf{Q}$ obtained from performing $n$ quantum measurements that cannot be reproduced by any no-signaling correlations with at most genuinely $(n-1)$-wise incompatible measurements, i.e., $ \mathbf{Q} \notin \mathrm{GNS}_{n-1} $.
\end{mainresult}
\begin{proof}
Let us start by analytically analyzing why the $M_{3322}$ inequality cannot be violated by distributions $\mathbf{P} \in  \mathrm{GNS}_{2}$. To make the proof as vivid as possible, we will work with the Collins-Gisin table notation of behaviours $\mathbf{P}$ and Bell inequalities to understand the structure of the considered inequalities better. In particular, any normalized no-signaling behaviour $\mathbf{P}$ of two-outcome measurements in uniquely characterized by a table (shown here for three settings)
\begin{align}
&\begin{array}{c||c c c}
 & P(A_{1}) & P(A_{2}) & P(A_{3}) \\
 \hline\hline
 P(B_{1}) & P(A_{1}B_{1}) & P(A_{2}B_{1}) & P(A_{3}B_{1})  \\
 P(B_{2}) & P(A_{1}B_{2}) & P(A_{2}B_{2}) & P(A_{3}B_{2}) \\
 P(B_{3}) & P(A_{1}B_{3}) & P(A_{2}B_{3}) & P(A_{3}B_{3})
\end{array},
\end{align}
where $P(A_{1}) \coloneqq p_A(a = 0 \vert x = 1)$, $ P(A_{1}B_{1}) \coloneqq p(a = 0, b = 0 \lvert x = 1, y=1)$ and similarly for the other terms with a straightforward generalization to more settings. Similarly, we represent Bell inequalities by a table of coefficients corresponding to the probability they are multiplied with. In this notation, the $M_{3322}$ inequality is given by
\begin{align}
M_{3322}=
&\begin{array}{c||c c c}
 & -2 & \phantom{-}0 &  \phantom{-}0  \\
 \hline\hline
 -2 & \phantom{-}1 & \phantom{-}1 & \phantom{-}1 \\
  -1 & \phantom{-}1 &  \phantom{-}1  & -1 \\
  \phantom{-} 0 & \phantom{-}1 &  -1  & \phantom{-}0 \\
\end{array} \leq 0.
\end{align}
Now, as we are restricting ourselves only to the no-signaling principle, consider a no-signaling distribution $ \mathbf{P} $ such that $ M_{3322} \odot \mathbf{P} > 0  $, where the symbol $ \odot $ denotes the application of the Bell functional onto the distribution $\mathbf{P}$. Note that for any no-signaling distribution the marginal distributions upper bound their involved correlations, i.e., $P(A_{k}) \geq P(A_{k}B_{j})$ and $P(B_{j}) \geq P(A_{k}B_{j})$. From this, we can deduce that 
\begin{align}
&[P(A_2 B_1) - P(B_1)] + [P(A_2B_2)-P(B_2)] \nonumber \\
&+[P(A_1 B_3)-P(A_1)] - P(A_2 B_3) \leq 0,   \label{Eq:ProofExample}
\end{align}
for any-no signaling distribution $ \mathbf{P} $. Note that we also used here that probabilities, such as the term $ P(A_2 B_3) $ are non-negative. Hence, it follows that $M_{3322} \odot \mathbf{P} > 0$ implies also $ M_{3322}^{(1,2)} \odot \mathbf{P} > 0 $, where $M_{3322}^{(1,2)}$ is given by
\begin{align}
M_{3322}^{(1,2)}=
&\begin{array}{c||c c c}
 & -1 & \phantom{-}0 &  \phantom{-}0  \\
 \hline\hline
 -1 & \phantom{-}1 & \phantom{-}0 & \phantom{-}1 \\
  \phantom{-} 0 & \phantom{-}1 &  \phantom{-}0  & -1 \\
  \phantom{-} 0 & \phantom{-}0 &  \phantom{-}0  & \phantom{-}0 \\
\end{array} \leq 0,
\end{align} 
as the term in Eq.~\eqref{Eq:ProofExample} only contributes non-positively towards the $M_{3322}$ inequality. Note that $M_{3322}^{(1,2)}$ is just the CHSH inequality involving Bob's measurements $(s = 1,t =2)$. By noting that other terms that only contribute negatively towards the Bell score of the $M_{3322}$ can be eliminated similarly, we also deduce that $M_{3322} \odot \mathbf{P} > 0$ implies a violation of the inequalities
\begin{align}
M_{3322}^{(1,3)} =
&\begin{array}{c||c c c}
 & -1 & \phantom{-}0 &  \phantom{-}0  \\
 \hline\hline
  -1 & \phantom{-}1 & \phantom{-}1 & \phantom{-}0 \\
  \phantom{-} 0 & \phantom{-}0 &  \phantom{-}0  & \phantom{-}0 \\
  \phantom{-} 0 & \phantom{-}1 &  -1  & \phantom{-}0 \\
\end{array} \leq 0,
\end{align} 
and 
\begin{align}
M_{3322}^{(2,3)} =
&\begin{array}{c||c c c}
 & -1 & \phantom{-}0 &  \phantom{-}0  \\
 \hline\hline
 \phantom{-}0 & \phantom{-}0 & \phantom{-}0 & \phantom{-}0 \\
  -1 &\phantom{-}1 &  \phantom{-}1  & \phantom{-}0 \\
  \phantom{-} 0 & \phantom{-}1 &  -1  & \phantom{-}0 \\
\end{array} \leq 0.
\end{align}
Now, this means every pair of two measurements of Bob has to lead to nonlocal correlations, as each of the above versions of the $M_{3322}^{(s,t)}$ is equivalent to the \ac{CHSH} inequality for the pair $(s,t)$. Hence, every pair of Bob's measurements has to be incompatible. By contra position, if one the pairs $(s,t)$ is compatible, it holds $M_{3322}^{(s,t)} \leq 0$. Consequently, it holds $M_{3322} \leq 0$ for all no-signaling distributions $\mathbf{P}$ in which some pair $(s,t)$ is compatible. The final question is, whether a behavior $ \mathbf{P} $ s.t. $M_{3322} \odot \mathbf{P}> 0$ could still be written as a convex combination of different $ \mathbf{P}^{(s,t)} $ according to Eq.~\eqref{DefnIncompDI}. However, using the fact that $M_{3322}$ is a linear in the probabilities and Eq.~\eqref{DefnIncompDI} is a convex decomposition, it follows directly that that $ M_{3322} \odot \mathbf{P} \leq 0 $ for any behavior $\mathbf{P} \in  \mathrm{GNS}_{2}$. \\
\indent For general $n \geq 2$, we consider the $M_{nn22}$ inequalities \cite{Brunner2006}, derived from slightly varying the family of $I_{nn22}$ inequalities \cite{Collins2004}. More formally, it holds $M_{nn22} \coloneqq I_{nn22} - (n-2)p_A(a = 0 \vert x = 1)$ or 
\begin{align}
M_{nn22}=
\begin{array}{c|| cccccc}
\mbox{} & -(n-1) & \phantom{-}0 & \phantom{-}0 & \cdots & \phantom{-}0 & \phantom{-}0  \\ \hline \hline
-(n-1) & \phantom{-}1 &  \phantom{-}1 &  \phantom{-}1 & \cdots & \phantom{-}1 & \phantom{-}1 \\
-(n-2) & \phantom{-}1 &  \phantom{-}1 &  \phantom{-}1 & \cdots & \phantom{-}1 & -1 \\
-(n-3) & \phantom{-}1 &  \phantom{-}1 &  \phantom{-}1 & \cdots & -1 & \phantom{-}0 \\
 \phantom{-}\vdots &  \phantom{-}\vdots & \mbox{} & \mbox{} & \mbox{} & \mbox{} &  \phantom{-}\vdots \\
-1 & \phantom{-}1 & \phantom{-}1 & -1 & \cdots & \phantom{-}0 & \phantom{-}0 \\
\phantom{-} 0 & \phantom{-}1 & -1 & \phantom{-}0 & \cdots & \phantom{-}0 & \phantom{-}0
\end{array} \leq 0. 
\end{align}
We show the full proof for general $n$, generalizing the above analysis in the Appendix. More specifically, we show that there exist a version of the \ac{CHSH} inequality for every pair $(s,t)$ of Bob's measurements that has to be violated if $M_{nn22}$ is violated. This means that a distribution $ \mathbf{P} $ that violates $M_{nn22}$ will exhibit nonlocality (and hence incompatibility) in each pair $ (s, s \neq t) $. Again, by the convexity of the problem and using the fact that $M_{nn22}$ is linear, it holds $M_{nn22} \odot \mathbf{P} \leq 0$ for all correlations $\mathbf{P} \in  \mathrm{GNS}_{n-1}$ originating from at most $(n-1)$-wise genuinely incompatible measurements on Bob's side according to Eq.~\eqref{DefnIncompDI}. \\
\indent 
Finally, it is know that all $I_{nn22}$ can be violated via the specific implementation found in \cite{PhysRevLett.104.060401}. It can be checked that the same holds for the $M_{nn22}$ inequalities. In particular, we use a quantum state 
\begin{align}
\lvert \Psi_{\epsilon} \rangle = \sqrt{\dfrac{1-\epsilon^2}{n-1}} \sum_{k = 1}^{n-1} \lvert k k \rangle + \epsilon \lvert nn \rangle,  \label{StateViolation}
\end{align} 
of local dimension $n$, parameterized by $ \epsilon \in [0,1] $. Using particular projective measurements defined in \cite{PhysRevLett.104.060401}, we obtain a Bell value of 
\begin{align}
M_{nn22}(\mathbf{Q}) =  \epsilon^2(q_0^2 n-(n-1)),    \label{Mnn22value}
\end{align}
where $\epsilon^2 = \tfrac{1-q_0^2}{1+[(n-1)^2-1]q_0^2}$ and $0 \leq q_0 \leq 1 $. This gives a violation which is strictly positive (but arbitrarily small for large $n$) whenever $ \sqrt{\tfrac{n-1}{n}} < q_0 < 1 $. This finishes the proof. 
\end{proof}
\indent \textit{Toward experimental implementations.}\textemdash As our proof provides specific Bell inequalities that can be violated by a specific (known) quantum implementation, it raises the question whether our results could experimentally be verified for a low number of settings $n$. Indeed, the $M_{3322}$ inequality has experimentally already been violated in \cite{PhysRevX.5.041052}, even with qubit systems. Meaning the first interesting case arises for $n = 4$. One challenge that an experimental verification has to face is the relatively low violation achievable according to Eq.~\eqref{Mnn22value}, resulting in high fidelity requirements. This seems to be a general feature of the family of $M_{nn22}$ inequalities and not particular to the specific implementation used here, as the maximal possible quantum violation of the $M_{nn22}$ inequality decreases generally for all quantum correlations, according to the third level of the NPA hierarchy \cite{PhysRevLett.98.010401}. For instance, while it holds in general quantum theory that $M_{3322}(\mathbf{Q}) \leq 0.0324$, already for $n=5$ measurements we only can hope to observe a violation of one order of magnitude less, i.e., $M_{5522}(\mathbf{Q}) \leq 0.0040$. In contrast to that, by maximizing Eq.~\eqref{Mnn22value} we can reach a value of $M_{3322}(\mathbf{Q}) \approx 0.0239 $ for $n=3$ measurements and $M_{5522}(\mathbf{Q}) \approx 0.0035$ for $n=5$ measurements. \\
\indent Analyzing whether this trend is specific to the family of $ M_{nn22} $ inequalities or a generic feature of the considered problem is generally hard, as it requires to find other families of Bell inequalities that are obeyed by all no-signaling distributions of at most genuinely $(n-1)$-wise incompatible measurements, which appears to be a hard task. Moreover, it is known that binary quantum correlations (in arbitrary dimensions) are quite restrictive in their strength, at least when traceless observables are considered \cite{PhysRevA.73.062105}. \\
\indent Nevertheless, recent experimental progress in high-dimensional detection loophole-free Bell tests \cite{PhysRevLett.129.060402} have been able to generate the quantum state in Eq.~\eqref{StateViolation} for $n=4$ with a fidelity of $99.5\%$ using path entanglement. While different measurements and a different Bell inequality have been considered in \cite{PhysRevLett.129.060402}, the employed techniques and measurements are close \cite{PhysRevLett.104.060401} to what is required for a violation of the $M_{4422}$ inequality using four dimensional systems. In conclusion, these recent developments still leave the question of an experimental violation of the $M_{4422}$ and hence an experimental falsification of no-signaling theories using only genuinely $3$ measurements open but within reach of what is accessible with state-of-the-art technologies. \\
\indent \textit{Discussion.}\textemdash We have proven that, for every finite integer $n \geq 2$, there exist quantum correlations that employ genuinely $n$-wise incompatible measurements in any no-signaling theory. That is, there cannot exist a no-signaling theory in which the predictions of quantum theory are reproduced by using only convex combinations of genuinely $(n-1)$-wise incompatible measurements across different partitions. If one would claim that such a theory exists, a violation of the corresponding $ M_{nn22} $ inequality would falsify said theory. Our proof relies on an analytic application of the framework of incompatibility structures to the family of $M_{nn22}$, showing that it is witness of genuine $n$-wise incompatibility when distributions $\mathbf{P}$ are only restricted by the no-signaling principle. We then argued that a practical verification of our results for low $n$, with particular focus on the case $n = 4$ are within experimental reach.  \\
\indent Our works proves that any no-signaling theory that reproduces the predictions of quantum theory does not only require more than $2$ measurements \cite{PhysRevLett.123.180401} but the number of required measurements is actually unbounded. At the same time, our results provide also an analogue to the findings in \cite{PhysRevLett.117.150401,PhysRevA.96.032104} (see also \cite{PhysRevLett.120.180402} for a corresponding experimental implementation), which imply that the predictions of quantum theory cannot be reproduced by a finite number of measurement outcomes in any no-signaling theory. There is an interesting parallel between the latter works and our results, as they require the Hilbert space dimension of the quantum system used for the Bell inequality violation to grow with $n$. While it is unclear whether this phenomenon is a necessity (in particular because $M_{3322}$ can be violated with qubits) it is reasonable to assume that it becomes increasingly harder to distinguish quantum theory from a no-signaling theory restricted to $(n-1)$ measurements, as the relative difference between $(n-1)$ and $n$ measurements shrinks. High-dimensional quantum systems might therefore be generally required to demonstrate this decreasing gap between the theories. \\
\indent Except the above question our work also initiates the search for
more suitable Bell inequalities and their quantum violation, such that experimental demands in terms of noise-robustness and detector efficiency could be lowered. It would also be interesting to see which other incompatibility structures \cite{PhysRevA.89.052126, Heinosaari2008,Liang2011,PhysRevLett.123.180401,PhysRevA.110.L060201} could be ruled out on the level of the no-signaling assumption only. Finally, it might be insightful to consider concepts for incompatibility and entanglement theory simultaneously to rule out possible post-quantum theories. \\

\begin{acknowledgments}
The author thanks Matthias Kleinmann, Marco Túlio Quintino, Marc-Olivier Renou, and Isadora Veeren for helpful discussions.
LT acknowledges funding from the ANR through the JCJC grant LINKS (ANR-23-CE47-0003).
\\[1em]
\end{acknowledgments}

\input{myacronyms}

\bibliography{bibliography_2.bib}

\newpage

\onecolumngrid

\section*{Appendix for "Quantum correlations cannot be reproduced with a finite number of measurements in any no-signaling theory"}

In this Appendix, we give a detailed proof of our main result from the main text. In order to make this document self-contained, we repeat the necessary concepts and the notation of the manuscript and provide aiding examples. \\
\indent We consider quantum measurements, which are described by positive semidefinite operators $A_{a \vert x} \geq 0$ acting on a Hilbert space $\mathcal{H} \cong \mathds{C}^{d}$ of dimension $d$, such that the completeness relation $\sum_{a} A_{a \vert x} = \mathds{1}$, (where $\mathds{1}$ is the identity operator), is fulfilled. Here, $a$ denotes the outcome corresponding to the measurement $x$. We say an assemblage $\mathcal{A} = \lbrace \lbrace A_{a \vert x} \rbrace_a \rbrace_x$ is \emph{jointly measurable} if it can be written such that 
\begin{align}
A_{a \vert x} = \sum_{\lambda} p(a \vert x, \lambda) G_{\lambda}, \ \forall \ a,x, \label{DefJMSM}  \end{align}
holds. Here, $ \lbrace G_{\lambda} \rbrace_{\lambda} $ is the so-called parent \ac{POVM} from which we simultaneously recover the statistics of the whole assemblage by post-processing with the probabilities $p(a \vert x \lambda)$. Eq.~\eqref{DefJMSM} can be understood as checking whether the assemblage $\mathcal{A}$ can be seen as one effective measurement (via the parent \ac{POVM}). However, if Eq.~\eqref{DefJMSM} cannot be satisfied for a given assemblage $\mathcal{A}$, it does not guarantee that $\mathcal{A}$ cannot be reduced to effectively two or more measurements. To do so, we have to make use of more general incompatibility structures and the notion of genuine $n$-wise incompatible assemblages. \\
\indent For a general assemblage $\mathcal{A}$, we say it is genuine $n$-wise incompatible (contains genuinely $n$ measurements) \cite{PhysRevLett.123.180401} if it cannot be decomposed such that 
\begin{align}
A_{a \vert x} = \sum_{(s,t \neq s) \in \mathcal{T}} p_{(s,t)}   F_{a \vert x}^{(s,t)}, \label{DefnGenuineIncompSM}
\end{align}
where $\mathcal{F}^{(s,t)} \in \mathrm{JM}^{(s,t)}$ are assemblages, employed with probability $p_{(s,t)}$, in which the pair $(s,t \neq s)$ is jointly measurable. This makes each $\mathcal{F}^{(s,t)}$ contain at most genuinely $(n-1)$-wise incompatible measurements (as the pair $(s,t \neq s)$ is one effective measurement) and the set $ \mathcal{T}$ contains all tuples $(s,t \neq s)$ constructed from the integers $\lbrace 1, \cdots, n \rbrace$. As an illustrative example, let us consider the incompatibility structures with respect to the three Pauli observables given by the Pauli matrices $X,Y,Z$, also used in \cite{PhysRevLett.123.180401}. More precisely, let $ \Pi_{a \vert x} $ with $a \in \lbrace \pm 1 \rbrace $ and $x \in \lbrace X,Y,Z \rbrace$ be the projectors corresponding to the Pauli observables. Moreover, consider a noise model of the form
\begin{align}
M^{\eta}_{a \vert x} = \eta   \Pi_{a \vert x} + (1-\eta) \dfrac{\mathds{1}}{2},  
\end{align}
with $ 0 \leq \eta \leq 1 $. It follows that all Pauli measurements are jointly measurable if $ \eta \leq \tfrac{1}{\sqrt{3}} $ and each pair of measurements is compatible for $ \eta \leq \tfrac{1}{\sqrt{2}} $. However, the Pauli measurements are only genuinely triplewise incompatible (can be regarded genuinely as $3$ measurements) if $ \eta > \tfrac{\sqrt{2}+1}{3} $. \\
\indent The restriction to the particular incompatibility structure in Eq.~\eqref{DefnGenuineIncompSM} leads to a restriction of the achievable behaviors $\mathbf{P}$. More precisely, let us assume a bipartite Bell scenario in which Bob's measurements obey the above incompatibility structure. It holds for the resulting correlations that 
\begin{align}
p(ab \vert xy) = \sum_{(s,t \neq s) \in \mathcal{T}} p_{(s,t)}  p^{(s,t)}(ab \vert xy),    \label{DefnGenuineIncompDISM}
\end{align}
where $\lbrace p^{(s,t)}(ab \vert xy) \rbrace_{a,b,x,y}$ describes a behavior $\mathbf{P}^{(s,t)}$ that is local once it is restricted to only the measurements $(s,t)$ on Bob's side. Certifying that $\mathbf{P}$ cannot be decomposed as above, proves in a device-independent way that Bob has genuinely access to $n$ measurements. Moreover, if we require $\mathbf{P}$ to only be a no-signaling distribution and not necessarily obey a quantum implementation, it is also a proof of the existence of genuine $n$-wise incompatible measurements on Bob's side in any no-signaling theory. \\
\indent Note that if we consider (linear) Bell inequalities, i.e., functionals of the form 
\begin{align}
\mathcal{S}(\mathbf{P}) = \sum_{a,b,x,y} C_{abxy} p(ab \vert xy) \leq L    
\end{align}
and we are interested in the maximum value $ \mathcal{S}^*(\mathbf{P}) $ over all no-signaling distributions $\mathbf{P}$ admitting a decomposition as in Eq.~\eqref{DefnGenuineIncompDISM}, it holds that 
\begin{align}
\mathcal{S}^*(\mathbf{P}) = \mathcal{S}_{max} = \max_{(s,t \neq s)}  \max_{\mathbf{P}^{(s,t)} }   \sum_{a,b,x,y} C_{abxy}   p^{(s,t)}(ab \vert xy), \label{Eq:Convexity}
\end{align}
by the convexity of the set of distributions admitting said decomposition. \\
\indent To prove our main result, we will use the Collins-Gisin table representation of Bell inequalities (and behaviors) \cite{Collins2004}. In the particular case of dichotomic (i.e., two-outcome) measurements, a behavior $\mathbf{P}$ can uniquely be defined (via normalization and no-signaling constraints) through a table of the form 
\begin{align}
&\begin{array}{c||c c c}
 & P(A_{1}) & P(A_{2}) & P(A_{3}) \\
 \hline\hline
 P(B_{1}) & P(A_{1}B_{1}) & P(A_{2}B_{1}) & P(A_{3}B_{1})  \\
 P(B_{2}) & P(A_{1}B_{2}) & P(A_{2}B_{2}) & P(A_{3}B_{2}) \\
 P(B_{3}) & P(A_{1}B_{3}) & P(A_{2}B_{3}) & P(A_{3}B_{3})
\end{array},
\end{align}
where $P(A_{1}) \coloneqq p_A(a = 0 \vert x = 1)$, $ P(A_{1}B_{1}) \coloneqq p(a = 0, b = 0 \lvert x = 1, y=1)$.
The generalization to more than three measurement settings is straightforward. Similarly, we can represent Bell-inequalities via their coefficients in a Collins-Gisin table. For instance, the \ac{CHSH} inequality \cite{PhysRevLett.23.880} (in Collins-Gisin notation) given by 
\begin{align}
p(00 \vert 11) + p(00 \vert 12) &+ p(00 \vert 21) - p(00 \vert 22) \nonumber \\
&-p_A(0 \vert 1) - p_B(0 \vert 1) \leq 0, \label{CHSHSM}   \end{align}
is represented by a table 
\begin{align}
\mathrm{CHSH}= &\begin{array}{c||c c}
 & -1 & \phantom{-}0  \\
 \hline\hline
 -1 & \phantom{-}1 & \phantom{-}1  \\
  \phantom{-}0 & \phantom{-}1 & -1  \\
\end{array} \leq 0.
\end{align}    
To compute the corresponding CHSH value of a behavior, we then write $ \mathrm{CHSH} \odot \mathbf{P}$, which corresponds to evaluating the CHSH functional for the distribution $\mathbf{P}$.
With this, we are in the position to state and then proof our main result. Let us denote the set of no-signaling correlations of at most genuinely $(n-1)$-wise incompatible measurements by $\mathrm{GNS}_{n-1}$. That is, $\mathbf{P} \in \mathrm{GNS}_{n-1}$ if a decomposition according to Eq.~\eqref{DefnGenuineIncompDISM} exists. 
\begin{mainresult} 
For every $n \geq 2$ there exist quantum correlations $\mathbf{Q}$ obtained from performing $n$ quantum measurements that cannot be reproduced by any no-signaling correlations with at most genuinely $(n-1)$-wise incompatible measurements, i.e., $ \mathbf{Q} \notin \mathrm{GNS}_{n-1} $.
\end{mainresult}
\begin{proof}
Consider the $M_{nn22}$ \cite{Brunner2006} inequality given by
\begin{align}
M_{nn22}=
\begin{array}{c|| cccccc}
\mbox{} & -(n-1) & \phantom{-}0 & \phantom{-}0 & \cdots & \phantom{-}0 & \phantom{-}0  \\ \hline \hline
-(n-1) & \phantom{-}1 &  \phantom{-}1 &  \phantom{-}1 & \cdots & \phantom{-}1 & \phantom{-}1 \\
-(n-2) & \phantom{-}1 &  \phantom{-}1 &  \phantom{-}1 & \cdots & \phantom{-}1 & -1 \\
-(n-3) & \phantom{-}1 &  \phantom{-}1 &  \phantom{-}1 & \cdots & -1 & \phantom{-}0 \\
 \phantom{-}\vdots &  \phantom{-}\vdots & \phantom{-}\vdots & \phantom{-}\vdots & \mbox{} & \phantom{-}\vdots &  \phantom{-}\vdots \\
-1 & \phantom{-}1 & \phantom{-}1 & -1 & \cdots & \phantom{-}0 & \phantom{-}0 \\
\phantom{-} 0 & \phantom{-}1 & -1 & \phantom{-}0 & \cdots & \phantom{-}0 & \phantom{-}0
\end{array} \leq 0. \label{Mn22}
\end{align}    
Let us first note that the $M_{nn22}$ inequality includes the $M_{(n-1)(n-1)22}$ inequality given by
\begin{align}
M_{(n-1)(n-1)22}=
\begin{array}{c|| cccccc}
\mbox{} & -(n-2) & \phantom{-}0 & \phantom{-}0 & \cdots & \phantom{-}0 & \phantom{-}0  \\ \hline \hline
-(n-2) & \phantom{-}1 &  \phantom{-}1 &  \phantom{-}1 & \cdots & \phantom{-}1 & \phantom{-}1 \\
-(n-3) & \phantom{-}1 &  \phantom{-}1 &  \phantom{-}1 & \cdots & \phantom{-}1 & -1 \\
 \phantom{-}\vdots &  \phantom{-}\vdots & \phantom{-}\vdots & \phantom{-}\vdots & \mbox{} & \phantom{-}\vdots &  \phantom{-}\vdots \\
-1 & \phantom{-}1 & \phantom{-}1 & -1 & \cdots & \phantom{-}0 & \phantom{-}0 \\
\phantom{-} 0 & \phantom{-}1 & -1 & \phantom{-}0 & \cdots & \phantom{-}0 & \phantom{-}0
\end{array} \leq 0.
\end{align} 
That is, if  $M_{nn22} \odot  \mathbf{P} > 0$ for some no-signaling behavior $\mathbf{P}$, it also follows that $M_{(n-1)(n-1)22}  \odot  \mathbf{P} > 0$.
This follows from the fact that it holds $P(A_{k}) \geq P(A_{k}B_{j})$ and $P(B_{j}) \geq P(A_{k}B_{j})$ for any no-signaling distribution, as marginal probabilities have to be at least as large as the correlation terms they appear in. With this, it can be directly seen that the first row of correlations in the Collins-Gisin table $M_{nn22}$ cannot exceed $(n-1)P(B_{1})+P(A_{1})$. Finally, the negative contribution of the entry corresponding to $p(a = 0,  b = 0 \vert x = n, y = 2)$ cannot help to violate the inequality. More formally, we use that it holds 
\begin{align}
\large[\sum_{j=2}^{n} P(A_j B_1) - (n-1)P(B_{1})\large]+[P(A_1 B_1)-P(A_1)] - P(A_nB_2)  \leq 0,   
\end{align}
for any no-signaling behavior $\mathbf{P}$. \\
\indent We will use this property as follows. We will first show that in order for the $ M_{nn22} $ inequality to be violated, it requires that Bob's first setting, i.e., $y=1$ violates the \ac{CHSH} inequality together with any other of his measurements $y \neq 1$. Now, the same has to be true for Bob's setting $y=2$, as it becomes the first setting of the $M_{(n-1)(n-1)22}$ inequality. Then, by induction, any of Bob's measurements $y=s$ has to be incompatible with any other setting $t \neq s$. Finally, by the same contra position argument as in the main text and using the linearity of the $M_{nn22}$ inequality together with fact that the decomposition in Eq.\eqref{DefnGenuineIncompDISM} is convex, we know that it holds  $M_{nn22} \odot \mathbf{R} \leq 0$ for any behavior $\mathbf{R} \in \mathrm{GNS}_{n-1}$ according to Eq.~\eqref{Eq:Convexity}.  \\
\indent To show that Bob's first measurement in the $ M_{nn22} $ inequality has to lead to nonlocal correlations (and hence be incompatible) with any other of the remaining $(n-1)$ measurements if only considered pairwise we proceed in the following way. We want to show that for measurement $y=1$ and every of the remaining measurements $y \neq 1$, a version of the \ac{CHSH} inequality 
\begin{align}
\mathrm{CHSH} = &\begin{array}{c||c c}
 & -1 & \phantom{-}0  \\
 \hline\hline
 -1 & \phantom{-}1 & \phantom{-}1  \\
  \phantom{-}0 & \phantom{-}1 & -1  \\
\end{array} \leq 0.
\end{align}   
has to be violated. This can be seen from counting arguments using the specific Collins-Gisin table representation of the $ M_{nn22} $ inequality in Eq.~\eqref{Mn22}. Note first that the anti-diagonal of $-1$ entries in the formulation of $M_{nn22} $ always ensures that we can find the corresponding coefficients for the correlation terms, i.e., the probabilities $p(ab \lvert xy) $ will appear with the same sign in the \ac{CHSH} inequality. Note further, that this always requires including Alice's first measurement, as this Alice's only measurement that contributes to the $  M_{nn22} $ with a negative marginal. For instance, for measurement $y = 1$ and $y = 2$, these correlation terms are given by the table
\begin{align}
\begin{array}{c|| cccccc}
\mbox{} & \phantom{-}0 & \phantom{-}0 & \phantom{-}0 & \cdots & \phantom{-}0 & \phantom{-}0  \\ \hline \hline
\phantom{-}0 & \phantom{-}1 &  \phantom{-}0 &  \phantom{-}0 & \cdots & \phantom{-}0 & \phantom{-}1 \\
\phantom{-}0 & \phantom{-}1 &  \phantom{-}0 &  \phantom{-}0 & \cdots & \phantom{-}0 & -1 \\
\phantom{-}0 & \phantom{-}0 &  \phantom{-}0 &  \phantom{-}0 & \cdots & \phantom{-}0 & \phantom{-}0 \\
 \phantom{-}\vdots &  \phantom{-}\vdots & \phantom{-}\vdots & \phantom{-}\vdots & \mbox{} & \phantom{-}\vdots &  \phantom{-}\vdots \\
\phantom{-}0 & \phantom{-}0 & \phantom{-}0 & \phantom{-}0 & \cdots & \phantom{-}0 & \phantom{-}0 \\
\phantom{-} 0 & \phantom{-}0 & \phantom{-}0 & \phantom{-}0 & \cdots & \phantom{-}0 & \phantom{-}0
\end{array}.
\end{align}
Of course, we also have to include Alice's and Bob's marginal corresponding to the \ac{CHSH} inequality. That is, we need to show that the inequality 
\begin{align}
\begin{array}{c|| cccccc}
\mbox{} & -1 & \phantom{-}0 & \phantom{-}0 & \cdots & \phantom{-}0 & \phantom{-}0  \\ \hline \hline
-1 & \phantom{-}1 &  \phantom{-}0 &  \phantom{-}0 & \cdots & \phantom{-}0 & \phantom{-}1 \\
\phantom{-}0 & \phantom{-}1 &  \phantom{-}0 &  \phantom{-}0 & \cdots & \phantom{-}0 & -1 \\
\phantom{-}0 & \phantom{-}0 &  \phantom{-}0 &  \phantom{-}0 & \cdots & \phantom{-}0 & \phantom{-}0 \\
 \phantom{-}\vdots &  \phantom{-}\vdots & \phantom{-}\vdots & \phantom{-}\vdots & \mbox{} & \phantom{-}\vdots &  \phantom{-}\vdots \\
\phantom{-}0 & \phantom{-}0 & \phantom{-}0 & \phantom{-}0 & \cdots & \phantom{-}0 & \phantom{-}0 \\
\phantom{-} 0 & \phantom{-}0 & \phantom{-}0 & \phantom{-}0 & \cdots & \phantom{-}0 & \phantom{-}0
\end{array} \leq 0,
\end{align}
can be extracted from the $M_{nn22} $ inequality by showing that the remaining terms will contribute non-positively towards the Bell value for any no-signaling distribution $ \mathbf{P} $. \\
Let us first list the following facts about the Collins-Gisin table corresponding to the $ M_{nn22} $ inequality:
\begin{enumerate}
    \item In row $r \in [1,n]$ there are $(n+1-r)$ positive ($+1$) valued correlation terms.

    \item In row $r$, there are $(n-r)$ negative ($-1$) marginals of Bob.

    \item In column $1$, there are $n$ positive ($+1$) correlation terms.

    \item In column $1$, there are $(n-1)$ negative ($-1$) marginals of Alice.  
\end{enumerate}
\indent With that, we proceed with the following algorithmic construction to show that $M_{nn22} \odot \mathbf{P} > 0$ implies a violation of the \ac{CHSH} inequality using Bob's measurements $y_1=1$ and $y_2 = s$ for every $s \in [2,n]$. For a given $s$, let us fix Alice's measurements $x_1=1$ and $x_2=n-s+2$. The following algorithm will extract the CHSH inequality for the settings $x_1,x_2,y_1,y_2$ while only removing non-positive terms for any no-signaling distribution $ \mathbf{P} $. 
\begin{enumerate}
    \item In row $y_1 = 1$, cancel out all $+1$ correlation terms except the term corresponding to $x_1$ and $x_2$, using $(n-2)$ of the negative marginals of Bob. This fixes the correct entries in the first row. That means, in the first row, the only remaining terms correspond to $-P(B_{y_1})$, $ P(A_{x_1} B_{y_1}), $ and $ P(A_{x_2} B_{y_1}) $. 

    \item In row $y_2 = s$, cancel out all $+1$ correlation terms except the term corresponding to $x_1 = 1$ using the $(n-s)$ negative marginal terms of Bob. This leaves only one remaining term, corresponding to $P(A_{x_1} B_{y_2})$.

    \item Set all $-1$ correlation terms except the term corresponding to $P(A_{x_2} B_{y_2})$ to $0$.

    \item In all rows $r$, except $y_1,y_2$, cancel out all $+1$ correlation terms except those corresponding to column $x_1 = 1$ using Bob's $(n-r)$ negative marginals.

    \item Cancel out all $+1$ correlation terms in column $x_1 = 1$, except those in row $y_1,y_2$, using $(n-2)$ of the negative marginal terms of Alice, leaving one term corresponding to $-P(A_{x_1})$ remaining. 
\end{enumerate}

It follows directly, that the above algorithm picks out the \ac{CHSH} inequality between the settings $x_1,x_2,y_1,y_2$, while only removing non-positive terms in each step, as marginals outweigh their corresponding correlation terms and probabilities are positive. Hence, we have shown the required property that none of Bob's measurements can be pairwise compatible when the $M_{nn22}$ inequality is violated. By the above reasoning, this directly implies $M_{nn22} \odot \mathbf{R} \leq 0$ for any no-signaling distribution $\mathbf{R} \in \mathrm{GNS}_{n-1}$. \\ 

\indent To finish the proof, we need to show that the $M_{nn22}$ inequality can be violated within quantum theory for every $n$. To show that this is indeed the case, we consider the family of states and measurements (defined for any $n$) given in \cite{PhysRevLett.104.060401}, which was developed to find a violation of the $I_{nn22}$ inequality for every $n$. As it turns out, the correlations are strong enough to also work in our case. For completeness, we repeat the construction of \cite{PhysRevLett.104.060401} here in the following. \\ 
\indent Before giving the explicit quantum strategy, we would like to point out the construction found in \cite{PhysRevLett.104.060401} is tailored to a different (but equivalent) representation of the $I_{nn22}$ (and therefore the resulting $M_{nn22}$) inequality then used in the remainder of this work. 
In particular, the quantum strategy down below can be used to violate the $\tilde{M}_{nn22}$ given by
\begin{align}
\tilde{M}_{nn22} = &- (n-1)  P(A_1) - \sum_{y = 2}^{n} P(B_y) + \sum_{y=1}^nP(A_1B_y) \\
&+\sum_{x=2}^nP(A_xB_x) - \sum_{1 \leq y < x \leq n} P(A_xB_y) \leq 0,    \nonumber
\end{align}
which in Collins-Gisin table notation is given by
\begin{align}
\tilde{M}_{nn22}=
\begin{array}{c|| ccccccc}
\mbox{} & -(n-1) & \phantom{-}0 & \phantom{-}0 & \phantom{-}0& \cdots & \phantom{-}0 & \phantom{-}0  \\ \hline \hline
\phantom{-}0 & \phantom{-}1 &  -1 &  -1 & -1 & \cdots & -1 & -1 \\
-1 & \phantom{-}1 &  \phantom{-}1 &  -1 & -1 &\cdots & -1 & -1 \\
-1 & \phantom{-}1 &  \phantom{-}0 &  \phantom{-}1 & -1&\cdots & -1 & -1 \\
-1 & \phantom{-}1 &  \phantom{-}0 &  \phantom{-}0 & \phantom{-}1&\cdots & -1 & -1 \\
 \phantom{-}\vdots &  \phantom{-}\vdots & \phantom{-}\vdots  & \phantom{-}\vdots & \mbox{} & \ddots & \mbox{} &  \phantom{-}\vdots \\
 -1 & \phantom{-}1 & \phantom{-}0 & \phantom{-}0 & \phantom{-}0 &\cdots &  \phantom{-}1 &-1 \\
-1 & \phantom{-}1 & \phantom{-}0 & \phantom{-}0 & \phantom{-}0 & \cdots &  \phantom{-}0 &\phantom{-}1 
\end{array} \leq 0. 
\end{align} 
However, the inequalities $ {M}_{nn22} \leq 0  $ and $ \tilde{M}_{nn22} \leq 0 $ are simply related by symmetries that do not alter local bounds or any achievable quantum violations. In particular, after exchange of Alice's settings (corresponding to exchanging columns)  $x \leftrightarrow n-x+2$ for all $x \geq 2$ and exchanging Alice's outcomes $ a = 0 \leftrightarrow a = 1 $ for all $x \geq 2$, which amounts to replacing $P(A_x B_y)$ with $P(B_y)-P(A_xB_y)$ and  $P(A_x)$ with $ 1 - P(A_x) $, we obtain the form of the ${M}_{nn22}$ inequality used in Eq.~\eqref{Mn22}. Hence, the following proof for a quantum violation of $\tilde{M}_{nn22} \leq 0$ translates directly to a quantum violation of $M_{nn22} \leq 0$ with the same Bell score. \\ 
\indent Consider a bipartite quantum state 
\begin{align}
\lvert \Psi_{\epsilon} \rangle = \sqrt{\dfrac{1-\epsilon^2}{n-1}} \sum_{k = 1}^{n-1} \lvert k k \rangle + \epsilon \lvert nn \rangle,  
\end{align} 
of local dimension $n$, parameterized by $ \epsilon \in [0,1] $. Furthermore, let $A_x = A_x^+ - A_x^-$, and  $B_y = B_y^+ - B_y^-$ be dichotomic observables for Alice and Bob respectively, where $A_x^{\pm}$ and $B_y^{\pm}$ are the corresponding projectors. Let us take the the projectors $A_x^+$ ($B_y^+$) to be rank-one projectors, i.e., $A_x^+ = \lvert a_x \rangle \langle a_x \rvert$ ($ B_y^+ = \lvert b_y \rangle \langle b_y \rvert $)
parameterized by a unit vector $\Vec{A}_x \in \mathds{R}^n$ ($\Vec{B}_y \in \mathds{R}^n$), such that $ \lvert a_x \rangle = \sum_{i = 1}^{n} \vec{A}_{xi} \lvert i \rangle $ ($ \lvert b_y \rangle = \sum_{i = 1}^{n} \vec{B}_{yi} \lvert i \rangle $). Moreover, let the $\Vec{A}_x$ be given as in \cite{PhysRevLett.104.060401}:
\begin{align}
&\vec{A}_1 = (0, \cdots, 0, 0, 0, 1) \\
&\vec{A}_2 = (0, \cdots, 0, -p_2, \dfrac{p_1}{n-1}, p_0) \nonumber \\
&\vec{A}_3 = (0, \cdots, -p_3, \dfrac{p_2}{n-2}, \dfrac{p_1}{n-1}, p_0) \nonumber \\
&\vdots \nonumber \\
&\vec{A}_{(n-1)} = (-p_{(n-1)}, \cdots, \dfrac{p_3}{n-3}, \dfrac{p_2}{n-2}, \dfrac{p_1}{n-1}, p_0) \nonumber \\
&\vec{A}_n = (p_{(n-1)}, \cdots, \dfrac{p_3}{n-3}, \dfrac{p_2}{n-2}, \dfrac{p_1}{n-1}, p_0), \nonumber 
\end{align}
where $ p_0^2 = \tfrac{1}{n} $, $ p_1^2 = \tfrac{n-1}{n}$, and $ p_{k+1}^2 = (1-\tfrac{1}{(n-k)^2})p_k^2 $ for $k \geq 1$. Similarly, the $\Vec{B}_y$ are given by
\begin{align}
&\vec{B}_1 = (0, \cdots, 0, 0, -q_1, q_0) \\
&\vec{B}_2 = (0, \cdots, 0, -q_2, \dfrac{q_1}{n-1}, q_0) \nonumber \\
&\vec{B}_3 = (0, \cdots, -q_3, \dfrac{q_2}{n-2}, \dfrac{q_1}{n-1}, q_0) \nonumber \\
&\vdots \nonumber \\
&\vec{B}_{(n-1)} = (-q_{(n-1)}, \cdots, \dfrac{q_3}{n-3}, \dfrac{q_2}{n-2}, \dfrac{q_1}{n-1}, q_0) \nonumber \\
&\vec{B}_n = (q_{(n-1)}, \cdots, \dfrac{q_3}{n-3}, \dfrac{q_2}{n-2}, \dfrac{q_1}{n-1}, q_0), \nonumber 
\end{align}
where $ q_0^2 + q_1^2 =1 $ and  $  q_{k+1}^2 = (1-\tfrac{1}{(n-k)^2})q_k^2 $ for $ k \geq 1 $. \\
\indent From the above state and measurements, it can be calculated that one finds for the relevant probabilities:
\begin{align}
&Q(A_{1}) \coloneqq p_A(a = 0 \vert x = 1) = \epsilon^2, \\ 
&Q(B_{y}) \coloneqq p_B(b = 0 \vert y) = \dfrac{1-\epsilon^2}{(n-1)}(1-q_0^2)+\epsilon^2q_0^2 \nonumber \\
&Q(A_{1}B_{y}) \coloneqq p(a = 0, b = 0 \vert x = 1, y) = \epsilon^2 q_0^2 \ \ \text{for} \ 1 \leq y \leq n \nonumber \\
&Q(A_{x}B_{x}) \coloneqq p(a = 0, b = 0 \vert x, x) = \Big(  \sqrt{\dfrac{1-\epsilon^2}{n-1}}p_1q_1+\epsilon p_0 q_0 \Big)^2 \ \text{for} \ x \geq 2 \nonumber \\
&Q(A_{x}B_{y}) \coloneqq p(a = 0, b = 0 \vert x, y) = \Big(  \sqrt{\dfrac{1-\epsilon^2}{n-1}}\dfrac{p_1q_1}{1-n}+\epsilon p_0 q_0 \Big)^2 \ \text{for} \ x > y \geq 1. \nonumber
\end{align}
\indent It follows that these correlations violate the $\tilde{M}_{nn22}$ inequality up to a value of
\begin{align}
\tilde{M}_{nn22}(\mathbf{Q}) = \epsilon^2(q_0^2 n-(n-1)),    
\end{align}
where $ \epsilon [0,1] $ is given such that
\begin{align}
\epsilon^2 = \dfrac{1-q_0^2}{1+[(n-1)^2-1]q_0^2},    
\end{align}
for some parameter $ 0 \leq q_0 \leq 1 $. This gives a violation which is strictly positive (but arbitrarily small for large $n$) whenever $ \sqrt{\tfrac{n-1}{n}} < q_0 < 1 $. This finishes the proof.

\end{proof}

\end{document}

%% file: myacronyms.tex
\begin{acronym}[CGLMP]\itemsep 1\baselineskip
\acro{AGF}{average gate fidelity}
\acro{AMA}{associated measurement assemblage}

\acro{BOG}{binned outcome generation}

\acro{CGLMP}{Collins-Gisin-Linden-Massar-Popescu}
\acro{CHSH}{Clauser-Horne-Shimony-Holt}
\acro{CP}{completely positive}
\acro{CPT}{completely positive and trace preserving}
\acro{CPTP}{completely positive and trace preserving}
\acro{CS}{compressed sensing} 

\acro{DFE}{direct fidelity estimation} 
\acro{DM}{dark matter}

\acro{GST}{gate set tomography}
\acro{GPT}{general probabilistic theory}
\acroplural{GPT}[GPTs]{general probabilistic theories}
\acro{GUE}{Gaussian unitary ensemble}

\acro{HOG}{heavy outcome generation}

\acro{JM}{jointly measurable}

\acro{LHS}{local hidden-state model}
\acro{LHV}{local hidden-variable model}
\acro{LOCC}{local operations and classical communication}

\acro{MBL}{many-body localization}
\acro{ML}{machine learning}
\acro{MLE}{maximum likelihood estimation}
\acro{MPO}{matrix product operator}
\acro{MPS}{matrix product state}
\acro{MUB}{mutually unbiased bases} 
\acro{MW}{micro wave}

\acro{NISQ}{noisy and intermediate scale quantum}

\acro{POVM}{positive operator valued measure}
\acro{PR}{Popescu-Rohrlich}
\acro{PVM}{projector-valued measure}

\acro{QAOA}{quantum approximate optimization algorithm}
\acro{QML}{quantum machine learning}
\acro{QMT}{measurement tomography}
\acro{QPT}{quantum process tomography}
\acro{QRT}{quantum resource theory}
\acroplural{QRT}[QRTs]{Quantum resource theories}

\acro{RDM}{reduced density matrix}

\acro{SDP}{semidefinite program}
\acro{SFE}{shadow fidelity estimation}
\acro{SIC}{symmetric, informationally complete}
\acro{SM}{Supplemental Material}
\acro{SPAM}{state preparation and measurement}

\acro{RB}{randomized benchmarking}
\acro{rf}{radio frequency}

\acro{TT}{tensor train}
\acro{TV}{total variation}

\acro{UI}{uninformative}

\acro{VQA}{variational quantum algorithm}

\acro{VQE}{variational quantum eigensolver}

\acro{WMA}{weighted measurement assemblage}

\acro{XEB}{cross-entropy benchmarking}

\end{acronym}